\documentclass[preprint,12pt]{article}

\usepackage{amssymb}
\usepackage{amsmath}
\usepackage{hyperref}

\begin{document}

\centerline{}

\centerline {\Large{\bf On the twofold Moutard transformation }}

\centerline{}

\centerline{\Large{\bf of the stationary Schr\"odinger equation }}

\centerline{}

\centerline{\Large{\bf with axial symmetry  }}

\centerline{}

\centerline{\bf {A. G. Kudryavtsev}}

\centerline{}

\centerline{Institute of Applied Mechanics,}

\centerline{Russian Academy of Sciences, Moscow 125040, Russia}

\centerline{kudryavtsev\_a\_g@mail.ru}

\begin{abstract}

The generalized Moutard transformation of the stationary axially symmetric 
Schr\"odinger equation is considered. It is shown that a superposition of two 
 Moutard transformations can provide new potentials for the eigenvalue problem. 
Examples of two - dimensional  potentials and exact solutions
for the stationary axially symmetric Schr\"odinger equation 
are obtained as an application of the  twofold Moutard transformation.

\end{abstract}

\centerline{PACS: 02.30.Jr, 02.30.Ik, 03.65.Ge}

%\begin{keyword}

%Darboux transformation \sep Nonlocal variables \sep Exactly
%solvable Schrodinger equations

%\PACS 03.65.Ge \sep 03.65.Fd \sep 02.30.Ik

%\end{keyword}

%\end{frontmatter}

\section{Introduction}

The stationary Schr\"odinger equation 
$\left( \Delta -{\it u}\left( x,y,z \right) \right) Y \left( x,y,z\right) =0$,
where $ \Delta$ is Laplace operator, 
is of greate interest because it describes different physical phenomena.
In the case ${\it u}=-E+v\left( x,y,z \right)$ this equation 
describes nonrelativistic quantum system with energy $E$ \cite
{Landau}, in the case ${\it u}= -{\omega} ^{2}/{c\left( x,y,z
\right)}^{2}$ equation describes an acoustic wave
 with temporal frequency $\omega$ in inhomogeneous media with
sound velocity $c$ \cite {Morse}.
A lot of physical environments have axial symmetry.
In the case of axial symmetry  the stationary Schr\"odinger equation 
 in spherical coordinates has the form
\begin{equation} \label{eq1}
\left({\frac {{\frac {\partial }{\partial r}} \left( {r}^{2}{\frac {
\partial }{\partial r}} \right) }{{r}^{2}}}+
{\frac {{\frac {\partial }{\partial \theta}} \left( \sin \left( \theta
 \right) {\frac {\partial }{\partial \theta}} \right) }{{r}^{2}\sin \left( \theta \right) }}
-u \left( r, \theta \right) \right) Y \left( r,\theta \right)
 = 0
\end{equation}

The importance of the existence of exactly solvable problems of quantum 
mechanics is indisputable. An exactly solvable model is understood as a model 
that allows one to construct exact solutions in explicit form. Such models 
are important not only for describing physical systems, but also serve as 
reliable initial approximations when constructing perturbation theory, 
and are also useful for testing numerical algorithms.
Therefore, the search for new exact solutions of the Schr\"odingerr equation
is an actual theme of scientific research, and various methods of solution 
are used \cite {Infeld1951} -  \cite {Gordillo2024}.

The useful tool for one - dimensional Schr\"odinger equation is
the Darboux transformation \cite {Matveev1991}. The Moutard 
transformation  \cite {Moutard1878}, \cite {Athorne 1991} 
is the generalization of the  Darboux transformation for the flat 
two - dimensional Schr\"odinger equation. In the papers 
\cite {Kudryavtsev2013},  \cite {Kudryavtsev2016} the nonlocal 
Darboux transformation of the two - dimensional
stationary Schr\"odinger equation in cartesian coordinates was 
considered and its relation to the Moutard transformation was 
established. In the paper \cite {Kudryavtsev2020} the nonlocal Darboux 
transformation of the stationary axially symmetric 
Schr\"odinger equation is considered and it is shown that a special case of 
the nonlocal Darboux transformation provides the generalization of 
the Moutard transformation. In the present paper eigenvalue problem 
for the axially symmetric Schr\"odinger equation is considered and new
two - dimensional  potentials and exact solutions 
are obtained as an application of the  generalized Moutard transformation.

\section{Generalized Moutard transformation in spherical coordinates}

As a result of the transition from cylindrical to 
spherical coordinates, the formulas for the generalized Moutard transformation 
from work \cite {Kudryavtsev2020} take the form
\begin{equation} \label{eq2}
{\frac {\partial }{\partial \theta}} \left( {\it \tilde {Y} } \left( r,
\theta \right) {\it Y_0} \left( r,\theta \right) \sin \left( \theta
 \right)  \right) -\sin \left( \theta \right) r \left( {\it Y_0}
 \left( r,\theta \right)  \right) ^{2}{\frac {\partial }{\partial r}}
 \left( {\frac {Y \left( r,\theta \right) }{{\it Y_0} \left( r,\theta
 \right) }} \right)=0
\end{equation}
\begin{equation} \label{eq3}
{\frac {\partial }{\partial r}} \left( {\it\tilde {Y} } \left( r,\theta
 \right) {\it Y_0} \left( r,\theta \right) \sin \left( \theta
 \right) r \right) +\sin \left( \theta \right)  \left( {\it Y_0}
 \left( r,\theta \right)  \right) ^{2}{\frac {\partial }{\partial 
\theta}} \left( {\frac {Y \left( r,\theta \right) }{{\it Y_0} \left( 
r,\theta \right) }} \right)=0 
\end{equation}
\begin{multline} \label{eq4}
\tilde {u} \left( r , \theta \right) = u\left( r , \theta \right) 
-{\frac {\partial ^{2}}{\partial {r}^{2}}}\ln  \left( \sin \left( 
\theta \right)  \left( {\it Y_0} \left( r,\theta \right)  \right) ^{2
} \right)
\\
 -{\frac {{\frac {\partial }{\partial r}}\ln  \left( \sin
 \left( \theta \right)  \left( {\it Y_0} \left( r,\theta \right) 
 \right) ^{2} \right) }{r}}-{\frac {{\frac {\partial ^{2}}{\partial {
\theta}^{2}}}\ln  \left( \sin \left( \theta \right)  \left( {\it Y_0}
 \left( r,\theta \right)  \right) ^{2} \right) }{{r}^{2}}}
\end{multline}
Here, $Y_0$ and $Y$ are the solutions of equation \eqref{eq1} with the initial
potential $u$. The function $ \tilde {Y}$
defined as a solution of a consistent system of equations
\eqref{eq2} and \eqref{eq3} is a solution of equation \eqref{eq1} with the new
potential $\tilde {u}$. 

Note that 
$Y=Y_0,  \tilde {Y}= {\frac {1}{{\it Y_0} \sin \left( \theta \right) r}}$ 
 is a simple example of solution of equations  \eqref{eq2} and  \eqref{eq3}.

In the papers \cite {Kudryavtsev2020}, \cite {Kudryavtsev2021} 
some examples of two-dimensional
potentials and exact solutions of the 
time-independent axially symmetric Schr\"odinger equation 
have been obtained on the basis of the
generalized Moutard transformation formulas. 
This examples illustrate that repeated
application of the generalized Moutard transformation
can lead to the potentials that are more interesting
for physical interpretation than the potentials obtained
by the first transformation. In particular, repeated application 
of the  generalized Moutard transformation is effective for obtaining 
potentials that do not have a singularity. Likewise for the two-dimensional
flat Schr\"odinger equation in cartesian coordinates, it was
shown in the paper \cite {Tsarev2008} 
that the  twofold application of the classical
Moutard transformation is efficient for obtaining
nonsingular potentials. Similarly, nonsingular potentials
in cylindrical coordinates can be effectively
obtained using the  twofold generalized Moutard transformation
 \cite {Kudryavtsev2020}, \cite {Kudryavtsev2021} .

When studying the Moutard transformations of the Schrödinger equation 
with axial symmetry, another interesting property of twofold transformations 
was discovered  \cite {Kudryavtsev2020}, 
namely the ability to obtain the same transformation 
for eigenfunctions with different eigenvalues. 
The fact is that for the eigenvalue problem $u = -k^2 + v \left( r,\theta \right) $ 
and with a single application of the Moutard transformation \eqref{eq4} 
using the selected eigenfunction $Y_0 \left( r,\theta,k \right)$ 
corresponding to the eigenvalue $k$, the new potential
has the form 
${\tilde {u}}=-k^2+{\tilde {v}} \left( r,\theta,k \right)$.
Due to ${\tilde {v}}$ contains a dependence on k, 
one obtains different transformations  
for eigenfunctions with different eigenvalues.
Similarly, the classical Moutard transformation in the 
two-dimensional flat case leads to 
different transformations  
for eigenfunctions with different eigenvalues. 
 Unlike the Moutard transformation, the Darboux transformation 
gives us the transformations of all eigenfunctions. 
This difference between the two-dimensional case and the one-dimensional 
case is apparently reflected in the well-known fact about 
integrability only at a selected energy level in the two-dimensional case 
(see, for example,  \cite {Veselov1984},  \cite {Krichever2019}).

Let us consider $u=-k^2$ and $Y_0=\sin \left( kr\cos \left( \theta \right)  \right) $
as a selected  eigenfunction. From  \eqref{eq4}  we obtain
\begin{equation} \label{eq5}
\tilde {u}  =
-{k}^{2}+2\,{\frac {{k}^{2}}{ \left( \sin \left( kr\cos \left( \theta
 \right)  \right)  \right) ^{2}}}+{\frac {1}{{r}^{2} \left( \sin
 \left( \theta \right)  \right) ^{2}}}
\end{equation}
This is an example of ${\tilde {v}} $ dependence on $k$.
To avoid this dependence we use twofold Moutard transformation.
As a result of the transition from cylindrical to 
spherical coordinates, the formulas for 
the superposition of two generalized Moutard transformations
from work \cite {Kudryavtsev2020} take the form
\begin{equation} \label{eq6}
{\tilde {\tilde {u}}}  =
u \left( r,\theta \right) -2\,{\frac {\partial ^{2}}{\partial {r}^{2}}
}\ln  \left( F \left( r,\theta \right)  \right) -2\,{\frac {{\frac {
\partial }{\partial r}}\ln  \left( F \left( r,\theta \right)  \right) 
}{r}}-2\,{\frac {{\frac {\partial ^{2}}{\partial {\theta}^{2}}}\ln 
 \left( F \left( r,\theta \right)  \right) }{{r}^{2}}}
\end{equation}
\begin{multline} \label{eq7}
{\frac {\partial }{\partial \theta}}F \left( r,\theta \right) =
\\
\sin \left( \theta \right) {r}^{2} \left(  \left( {\frac {\partial }{
\partial r}}{\it Y_2} \left( r,\theta \right)  \right) {\it Y_1} \left( 
r,\theta \right) - \left( {\frac {\partial }{\partial r}}{\it Y_1}
 \left( r,\theta \right)  \right) {\it Y_2} \left( r,\theta \right) 
 \right) 
\end{multline}
\begin{multline} \label{eq8}
{\frac {\partial }{\partial r}}F \left( r,\theta \right) =
\\
-\sin \left( \theta \right)  \left(  \left( {\frac {\partial }{\partial 
\theta}}{\it Y_2} \left( r,\theta \right)  \right) {\it Y_1} \left( r,
\theta \right) - \left( {\frac {\partial }{\partial \theta}}{\it Y_1}
 \left( r,\theta \right)  \right) {\it Y_2} \left( r,\theta \right) 
 \right) 
\end{multline}
Here, $Y_1$ and $Y_2$ are the solutions of equation \eqref{eq1} 
with the initial potential $u$. 

Equations \eqref{eq7}, \eqref{eq8} are invariant under the substitution
${\it Y_1} \rightarrow {\it Y_2}, {\it Y_2} \rightarrow {\it Y_1}, F  \rightarrow -F$,
which represents the
commutativity of the generalized Moutard transformations.
The result does not depend on the order of
choice of the functions ${\it Y_1}, {\it Y_2}$ for the transformation.
We also note that $F$ is determined up to multiplication by a constant.
Simple examples of solutions for the equation  \eqref{eq1} with potential  \eqref{eq6} 
can be obtained by the formulas
$ {\tilde {\tilde {Y}}}_{1} = {\it Y_1} F^{-1},\,  
{\tilde {\tilde {Y}}}_{2} = {\it Y_2}F^{-1}$.

\section{Examples of  potentials and exact solutions}

Let us consider 
$u=-k^2, Y_1=\sin \left( kr\cos \left( \theta \right)  \right), 
Y_2=\cos \left( kr\cos \left( \theta \right)  \right) $.
From equations \eqref{eq7}, \eqref{eq8} we obtain
$F={r}^{2} \left( \sin \left( \theta \right)  \right) ^{2}+C$
where $C$ is an arbitrary constant.
 From formula \eqref{eq6} we get new potential
\begin{equation} \label{eq9}
{\tilde {\tilde {u}}}  =
-{k}^{2}+4\,{\frac {{r}^{2} \left( \sin \left( \theta \right)  \right) ^{2}
-C}{ \left( {r}^{2} \left( \sin \left( \theta \right)  \right) ^{2}+C
 \right) ^{2}}}
\end{equation}
Here ${\tilde {\tilde {v}}}$ does not depend on $k$ and we
have obtained new eigenvalue problem.
As a solution to the original eigenvalue problem with $u=-k^2$  
we will consider a plane wave ${{\rm e}^{ikr\cos \left( \theta \right) }}$.
Carrying out a  twofold Moutard transformation or using a simple formula 
$ {\it Y_2}F^{-1} + i  {\it Y_1}F^{-1}$ 
we obtain the following solution to equation  \eqref{eq1} 
with potential  \eqref{eq9}
\begin{equation} \label{eq10}
{\frac {{{\rm e}^{ikr\cos \left( \theta \right) }}}{{r}^{2} \left( \sin
 \left( \theta \right)  \right) ^{2}+C}} 
\end{equation}

As another example, consider two solutions  to equation
\eqref{eq1} with potential $u=-k^2$
\begin{equation} \label{eq11}
 Y_1=
{\frac {{{\sl J}_{p+1/2}\left(kr\right)}{\it P} \left( p,\cos
 \left( \theta \right)  \right) }{\sqrt {r}}}
\end{equation}
\begin{equation} \label{eq12}
Y_2=
{\frac {{{\sl Y}_{p+1/2}\left(kr\right)}{\it P} \left( p,\cos
 \left( \theta \right)  \right) }{\sqrt {r}}}
\end{equation}
where $p$ is a parameter, ${\sl J}_{p+1/2}, {\sl Y}_{p+1/2} $ are 
Bessel functions of the first and second kind, ${\it P}$ 
is Legendre function of the first kind. 
From equations \eqref{eq7}, \eqref{eq8} we obtain
$F$ depending only on $\theta$
\begin{equation} \label{eq13}
F_p \left( \theta \right) =
-\int \!\sin \left( \theta \right)  \left( {\it P} \left( p,\cos
 \left( \theta \right)  \right)  \right) ^{2}\,{\rm d}\theta
\end{equation}
 From formula \eqref{eq6} we get new potential
\begin{equation} \label{eq14}
{\tilde {\tilde {u_p}}}  =
-k^ 2 -2\,{\frac {{\frac {\partial ^{2}}{\partial {\theta}^{2}}}\ln 
 \left( F_p \left( \theta \right)  \right) }{{r}^{2}}}
\end{equation}
where $ {\tilde {\tilde {v}}} $ does not depend on $k$.
Note that $ {\tilde {\tilde {v}}}={\frac { f \left( \theta \right)  }{{r}^{2}}} $ 
 has the form of a Calogero-Moser potential. The problem of integrability of 
Calogero-Moser type potentials continues to attract the attention of 
mathematicians and physicists  \cite {Berest2023}.

For non-negative integer values of the parameter $p$, the function $F$ 
can be obtained explicitly. So for p=0,1,2 we have
 
$F_0=\cos \left( \theta \right) +C, F_1=\left( \cos \left( \theta \right)  \right) ^{3}+C,$

$F_2=\left( 9\, \left( \cos \left( \theta \right)  \right) ^{4}-10\,
 \left( \cos \left( \theta \right)  \right) ^{2}+5 \right) \cos
 \left( \theta \right) +C
$ 

where arbitrary constants $C$ for each $F_p$ can of course 
be chosen independently.

 From  \eqref{eq14}  we obtain the corresponding potentials
\begin{equation} \label{eq15}
{\tilde {\tilde {u}}}_{0} =
-{k}^{2}+2\,{\frac {\cos \left( \theta \right) C+1}{ \left( \cos
 \left( \theta \right) +C \right) ^{2}{r}^{2}}}
\end{equation}
\begin{equation} \label{eq16}
{\tilde {\tilde {u}}}_{1} =
-{k}^{2}+6\,{\frac {\cos \left( \theta \right)  \left(  \left( 3\,
 \left( \cos \left( \theta \right)  \right) ^{2}-2 \right) C+ \left( 
\cos \left( \theta \right)  \right) ^{3} \right) }{ \left(  \left( 
\cos \left( \theta \right)  \right) ^{3}+C \right) ^{2}{r}^{2}}}
\end{equation}
\begin{equation} \label{eq17}
{\tilde {\tilde {u}}}_{2} =
-{k}^{2} +10\,{\frac {N \left( \theta \right) }{\left(  \left( 9\,
 \left( \cos \left( \theta \right)  \right) ^{4}-10\, \left( \cos
 \left( \theta \right)  \right) ^{2}+5 \right) \cos \left( \theta
 \right) +C \right) ^{2}{r}^{2}}}
\end{equation}
where
\begin{multline*}
N \left( \theta \right)=
\left( 45\, \left( \cos \left( \theta \right)  \right) ^{4}-54\,
 \left( \cos \left( \theta \right)  \right) ^{2}+13 \right) \cos
 \left( \theta \right) C
\\
+ \left( 9\, \left( \cos \left( \theta
 \right)  \right) ^{4}+72\, \left( \cos \left( \theta \right) 
 \right) ^{2}-70 \right)  \left( \cos \left( \theta \right)  \right) ^
{4}+5
\end{multline*}

As examples of exact solutions to the scattering problem with potentials 
 \eqref{eq15} - \eqref{eq17}, we again consider a plane wave 
 ${{\rm e}^{ikr\cos \left( \theta \right) }}$ 
for the original $u=-k^2$  and carry out a twofold Moutard transformation 
with the corresponding $Y_1$ and $Y_2$ from formulas  \eqref{eq11} and 
 \eqref{eq12}. As a result, we obtain the following exact solutions
\begin{equation} \label{eq18}
{\tilde {\tilde {Y}}}_{0}  =
{{\rm e}^{ikr\cos \left( \theta \right) }} \left( 1+{\frac {i}{kr
 \left( \cos \left( \theta \right) +C \right) }} \right) 
\end{equation}
\begin{equation} \label{eq19}
{\tilde {\tilde {Y}}}_{1}  =
{{\rm e}^{ikr\cos \left( \theta \right) }} \left( 1+3\,{\frac {\cos
 \left( \theta \right)  \left( ikr\cos \left( \theta \right) -1
 \right) }{{k}^{2}{r}^{2} \left(  \left( \cos \left( \theta \right) 
 \right) ^{3}+C \right) }} \right)  
\end{equation}
\begin{multline} \label{eq20}
{\tilde {\tilde {Y}}}_{2}  =
\\
{{\rm e}^{ikr\cos \left( \theta \right) }} \left( 1+{\frac {M \left( r
,\theta,k \right) }{{r}^{3}{k}^{3} \left(  \left( 9\, \left( \cos
 \left( \theta \right)  \right) ^{4}-10\, \left( \cos \left( \theta
 \right)  \right) ^{2}+5 \right) \cos \left( \theta \right) +C
 \right) }} \right) 
\end{multline}
where
\begin{equation*}
M \left( r,\theta,k \right)   =
5\,i \left( 3\, \left( \cos \left( \theta \right)  \right) ^{2}-1
 \right)  \left(  \left( 3\, \left( \cos \left( \theta \right) 
 \right) ^{2}-1 \right) {r}^{2}{k}^{2}+6\,ikr\cos \left( \theta
 \right) -6 \right). 
\end{equation*}

\section{Results and discussion}

The eigenfunction of the eigenvalue problem of the two-dimensional 
Schr\"odinger operator allows us to obtain the Moutard transformation 
for the potential. However, the resulting potential depends on the 
eigenvalue corresponding to the eigenfunction. The article shows 
that in a number of cases, applying the Moutard transformation twice makes it 
possible to obtain a potential that does not depend on the eigenvalues. 
It is important to note that the Moutard transformation to obtain new 
solutions requires quadratures, so obtaining eigenfunctions in explicit 
analytical form for a new potential is not always possible. If for the 
new potential the eigenfunctions are obtained in explicit form, then 
we can talk about a new solvable two-dimensional eigenvalue problem. 
Here we say that a problem is solvable if it allows obtaining 
of solutions in an explicit functional form. This article provides examples 
of new solvable eigenvalue problems in the case of the Schr\"odinger operator with 
axial symmetry. For a new solvable problem, one can try again
to construct a twofold Moutard transformation. 
Finding all solvable eigenvalue problems of the Schrödinger operator 
that can be obtained using the twofold Moutard transformation remains 
an open problem.

\end{document}